# The leaky integrator with recurrent inhibition as a predictor

Revised 7/27/2016 16:02:00

Henning U. Voss

*Weill Cornell Medicine - Medical College, Citigroup Biomedical Imaging Center*
*516 East 72nd Street, New York, NY10021, USA.*

**Abstract:** It is shown that the leaky integrator, the basis for many neuronal models, possesses a negative group delay when a time-delayed recurrent inhibition is added to it. By means of this negative group delay, the leaky integrator becomes a predictor for some frequency components of the input signal. The prediction properties are derived analytically and an application to a local field potential is provided.

In "How delays affect neural dynamics and learning" [1], the authors state that "Integration and communication delays are ubiquitous, both in biological and man-made neural systems […] Indeed, delays should be considered as an additional media through which evolution, or skilled engineers, can achieve particular dynamical effects." In fact, time delays have an impact on the dynamics of neuronal networks, for example by causing oscillations and waves [2]. In this Note I would like to point out how time delays added to leaky integrators are defining predictors for smooth input signals. The underlying mechanism, negative group delay (NGD), to the best of my knowledge has not been used in the neurosciences so far.

The basic model is a leaky integrator with a recurrent, time-delayed inhibition. The leaky integrator is defined as usual as a capacitance, the integrator, in parallel to a resistance, the leak [3, 4]. The recurrent inhibition is modelled as a linear time-delayed feedback term with negative gain [5]. The leaky integrator with recurrent inhibition follows as

$$\dot{y}(t) = -a\, y(t) + b\, x(t) - c\, y(t - \tau)\,, \qquad (1)$$

where $a \geq 0$ is the leakage coefficient, $x(t)$ the input signal (zero-mean, generated by a stationary process), $b > 0$ the input scaling, $c \geq 0$ the (inhibitory) feedback gain, and $\tau > 0$ a time delay. For $c = 0$, Eq. (1) would simply be a leaky integrator, but for $c > 0$ it has an inhibitory feedback that enters the model as a delayed leak. Therefore, model (1) is referred to as a "delayed-leak integrator" (DLI). For $a = 0$ and $c > 0$, it describes a pure DLI without a conventional leak, which would have similar properties as the DLI with $a > 0$ but is not further considered here.

Equation (1) is linear and thus can be described by its frequency response function

$$H(\omega) = \frac{b}{a + i\omega + ce^{-i\omega\tau}} = \frac{b}{\beta}(R + iI), \qquad (2)$$

with $R = a + c\cos(\omega\tau)$, $I = c\sin(\omega\tau) - \omega$, $\beta = R^2 + I^2$ [6, 7]. It defines the steady-state input-output relationship between $x$ and $y$ in Fourier space as $Y(\omega) = H(\omega)X(\omega)$, where $f$ is frequency, $\omega = 2\pi f$, $x(t) = \int X(\omega)e^{i\omega t}d\omega$, and $y(t) = \int Y(\omega)e^{i\omega t}d\omega$. If written as $H(\omega) = G(\omega)e^{i\Phi(\omega)}$, its gain is $G(\omega) = b/\sqrt{\beta}$, and its phase is $\Phi(\omega) = \arg(R + iI)$. The frequency dependent group delay is

$$\delta(\omega) = -\frac{d\Phi(\omega)}{d\omega} = \frac{c\cos(\omega\tau) - c^2\tau - a(c\tau\cos(\omega\tau) - 1) + c\tau\omega\sin(\omega\tau)}{\beta}. \qquad (3)$$

It can be positive, zero, or negative. Negative group delay in general means a group advance, or real-time prediction of the input signal [8, 9]. To characterize the group delay for low frequency signal components, $\delta(\omega)$ is expanded for small $\omega$. Neglecting quadratic and higher order terms in the counter and denominator of the expansion, it follows

$$\delta_{\text{small }\omega} \approx \frac{1 - c\tau}{a + c}. \qquad (4)$$

This result has two important consequences for input signal components with small $\omega$:
(i) For $c\tau > 1$, the group delay is *negative*, a necessary condition for prediction.
(ii) The group delay is approximately *independent of $\omega$*, a necessary condition for distortion-free signal transfer [10].

These ideas were applied to a local field potential (LFP) from the left hippocampus (CA1) of a rat. The data consisted of the first 80 s of the "hc-5" set from CRCNS.org [11]. The input $x$ was defined as the average over all electrodes, normalized, and slightly lowpass filtered (cutoff at 27 Hz). The parameters $a = 2.00$ ms$^{-1}$ and $c = 1.40$ ms$^{-1}$ were estimated from a fit to the first 5 s of the data set with a simplex search algorithm and then used to model 16 contiguous intervals of 5 s each ($b = 0.6\,a$). Equation (1) was solved with a Runge-Kutta scheme with $\tau = 40.0$ ms and $y(t) = 0$ for $t \in [-\tau, 0]$. All computations were performed with MATLAB R2015a (The MathWorks, Inc., Natick, MA).

In Figure (a) 1.4 s of the input $x$ and corresponding output $y$ are shown. *It is evident that the DLI output $y$ at time $t$ (red) predicts the LFP input $x$ at a time $t + |\delta|$ (black) on average.* The group delay $\delta_{\text{small }\omega}$ is -16.2 ms (Figure (b), red dashed line shows $10\delta_{\text{small }\omega}$). Figure (b) also depicts a section of the estimated and analytic phase and gain of the frequency response function, including the first interval with NGD. Therefore, frequency components of $x$ within this interval are predicted by $y$. More specifically, the cross-correlation function (CCF) between $x$ and $y$ (Figure (c)) has a global maximum of 0.81 at $\delta$ = -7.2 ms. This result is reproducible: Out of the 16 data sections, 11 yielded a CCF with a global maximum between $\delta$ = -7.2 and -8.8 ms. The importance of signal frequency content is demonstrated in Figure (d); spectral components near the resonance of the frequency response function are amplified and become detrimental to prediction if dominating the signal.

A discussion concludes this Note: The real-time prediction of the LFP does not violate causality but follows from the delay-induced NGD [7] of the DLI (1). Prediction performance depends on model parameters and spectral properties of the data. For improper conditions the DLI might not predict or cause oscillatory instabilities [12]. Very recently it has been emphasized in this journal that anticipatory systems can defy the inference of the direction of information flow from data [13]. It would be interesting if this holds true for NGD systems, too [14-16]. DLIs might augment the related concept of neuronal anticipatory synchronization [17-20], which recently has been used to explain observations in brain dynamics [21]. Note that DLIs do not require a memory of past signal values, only of past predicted, already internalized, states, as Eq. (1) does not contain delayed inputs. It would be worth investigating how NGD systems fit into general theories of prediction [22] or how they perform as predictors in artificial neuronal networks. Since delay-induced NGD does not depend on a specific model for the signal, it is quite conceivable that biological neuronal networks might utilize this simple mechanism for real-time prediction, such that Baldi and Atiya's insights could be corroborated once again.

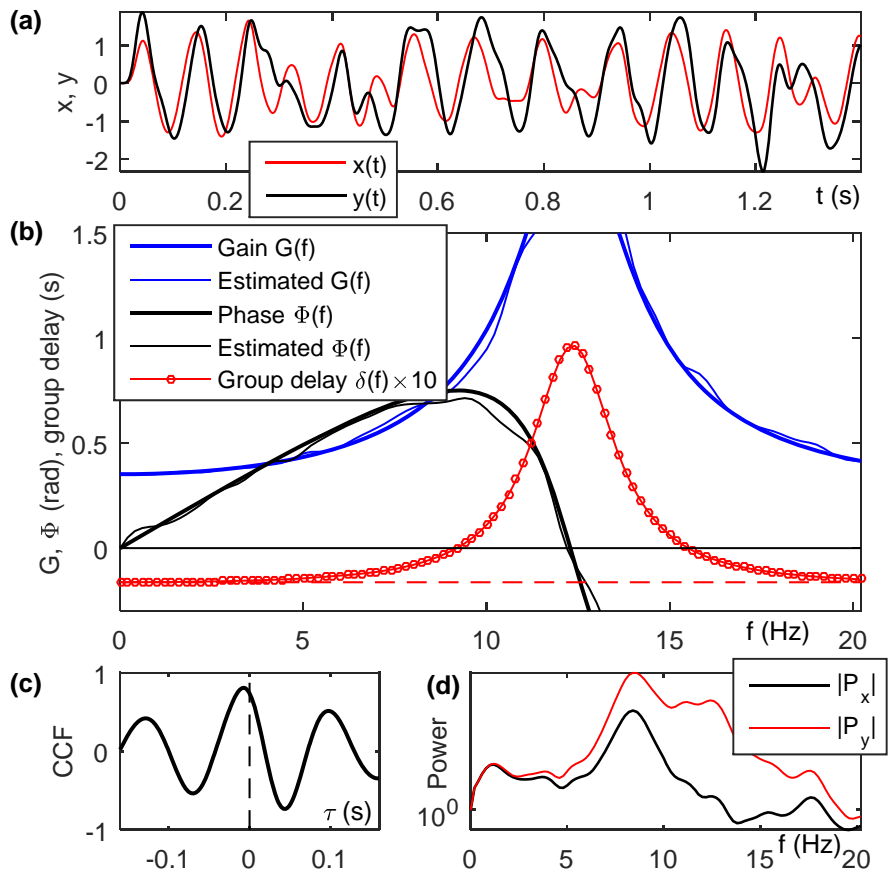

**Figure: Simulation of the DLI system with experimental input data.** Please refer to text for detailed description.